\documentstyle[11pt,aaspp4]{article}
\def \half {{\textstyle\frac{1}{2}}}

\def \kms {{\rm km~s^{-1}}}
\def \kpc {{\rm kpc}}
\def \hkpc {{h^{-1} \kpc}}
\def \mpc {{\rm Mpc}}
\def \hmpc {{h^{-1} \mpc}}

\def \ie {{i{.}e{.}}, }

\def \Om {{\Omega}}

\def \xb {{\bf{x}}}

\def \rb {{\bf{r}}}
\def \vb {{\bf{v}}}
\def \ub {{\bf{u}}}
\def \ab {{\bf{a}}}
\def \gb {{\bf{g}}}
\def \sb {{\bf{s}}}
\def \qb {{\bf{q}}}

\def \sP {{\cal P}}
\def \sS {{\cal S}}
\def \icP {{i \! \times \! \sP}}
\def \PcP {{\sP \! \times \! \sP}}
\def \ScP {{\sS \! \times \! \sP}}
\def \SncP {{\sS_n \! \times \! \sP}}

\def \sv {{\sigma_v}}
\def \svsq {{\sigma_v^2}}
\def \sz {{\sigma_z}}
\def \szsq {{\sigma_z^2}}

\def \rdotv {{\rb \!\cdot\! \vb}}
\def \rdota {{\rb \!\cdot\! \ab}}
\def \rdots {{\rb \!\cdot\! \sb}}

\def \rd {{\dot{r}}}
\def \rdd {{\ddot{r}}}

\def \avg#1{\langle #1 \rangle}

\def \rdsqavg {{\avg{\rd^2}}}
\def \rddavg {{\avg{\rdd}}}

\def \vsqavg {{\avg{v^2}}}

\def \vicPsqavg {{\avg{v_{\icP}^2}}}
\def \vScPsqavg {{\avg{v_{\ScP}^2}}}
\def \vSncPsqavg {{\avg{v_{\SncP}^2}}}

\def \vrsqavg {{\avg{v_r^2}}}
\def \vtsqavg {{\avg{v_t^2}}}
\def \vrvtsqavg {{\avg{2 v_r^2 - v_t^2}}}
\def \rdotaavg {{\avg{\rdota}}}
\def \rdotgavg {{\avg{\rb \!\cdot\! (\gb_j - \gb_i)}}}
\def \gbiavg {{\avg{\gb_i}}}
\def \gbjavg {{\avg{\gb_j}}}

\def \nub {{\bar{\nu}}}
\def \nb {{\bar{n}}}

\def \xo {{(1 + \xi)}}

\def \xidd {{\ddot{\xi}}}

\def \pr {{\partial_r}}

\def \pdd {{\partial_t^2}}

\def \dr {{\delta r}}
\def \dt {{\delta t}}

\def \Nd {{\dot{N}}}
\def \Ndd {{\ddot{N}}}

\def \BDM {\begin{displaymath}}
\def \EDM {\end{displaymath}}
\def \BEQ {\begin{equation}}
\def \EEQ {\end{equation}}
\def \BEQA {\begin{eqnarray}}
\def \EEQA {\end{eqnarray}}
\def \NN {\nonumber}
\def \BL {\begin{list}}
\def \EL {\end{list}}
\def \BENUM {\begin{enumerate}}
\def \EENUM {\end{enumerate}}
\def \BITEM {\begin{itemize}}
\def \EITEM {\end{itemize}}
\def \BARR {\begin{array}}
\def \EARR {\end{array}}
\def \BFIG {\begin{figure}}
\def \EFIG {\end{figure}}

\lefthead{KEPNER, SUMMERS, \& STRAUSS}
\righthead{Redshift Dispersion}

\begin{document}

\title{A New Statistic for Redshift Surveys: \\
the Redshift Dispersion of Galaxies}
\author{Jeremy~V.~Kepner\altaffilmark{1}, F~J~Summers\altaffilmark{2}, and
Michael~A.~Strauss\altaffilmark{3}}
\affil{Princeton University Observatory, Peyton Hall, Ivy Lane,
Princeton, NJ 08544--1001 \\
(jvkepner/strauss)@astro.princeton.edu, summers@astro.columbia.edu}
\authoremail{(jvkepner/strauss)@astro.princeton.edu, summers@astro.columbia.edu}
\altaffiltext{1}{DoE Computational Science Fellow}
\altaffiltext{2}{Current address: Columbia Astrophysics Lab, Mail Code
5247, 550 W 120th St, New York, NY 10027}
\altaffiltext{3}{Alfred P. Sloan Foundation Fellow}

%%%%%%%%%%%%%%%%%%%%%%%%%%%%%%%%%%%%%%%%%%%%%%%%%%%%%%%%%%%%
\begin{abstract}

  We present a new statistic---the redshift dispersion--- which may
prove useful for comparing next generation redshift surveys (e.g., the
Sloan Digital Sky Survey) and cosmological simulations. Our statistic is
specifically designed for the projection of phase space which is
directly measured by redshift surveys.  We find that the redshift
dispersion of galaxies as a function of the projected overdensity has a
strong dependence on the cosmological density parameter $\Omega$.  The
redshift dispersion statistic is easy to compute and can be motivated by
applying the Cosmic Virial Theorem to subsets of galaxies with the same
local density. We show that the velocity dispersion of particles in
these subsets is proportional to the product of $\Omega$ and the local
density.  Low resolution N-body simulations of several cosmological
models (open/closed CDM, CDM+$\Lambda$, HDM) indicate that the
proportionality between velocity dispersion, local density and $\Omega$
holds over redshift scales in the range 50 $\kms$ to 500 $\kms$.  The
redshift dispersion may provide an interesting means for comparing
volume-limited subsamples of the Sloan Digital Sky Survey to equivalent
N-body/hydrodynamics simulations.

\end{abstract}

%%%%%%%%%%%%%%%%%%%%%%%%%%%%%%%%%%%%%%%%%%%%%%%%%%%%%%%%%%%%


%%%%%%%%%%%%%%%%%%%%%%%%%%%%%%%%%%%%%%%%%%%%%%%%%%%%%%%%%%%%
\section{Introduction}
%%%%%%%%%%%%%%%%%%%%%%%%%%%%%%%%%%%%%%%%%%%%%%%%%%%%%%%%%%%%

  Many statistical measures have been developed to distinguish between
the various cosmological models, ranging from direct  measures of the
power spectrum and correlation function from redshift surveys, to
measures of velocity dispersion, bulk flows, and Mach number from
peculiar velocity surveys (cf., \cite{Strauss95} for a review). Each of
these measures is designed  to be sensitive to different aspects of
various models, such as $\Omega$, the initial power spectrum, or the
Gaussian character of the initial phase distribution.

  One can get an intuitive understanding of the dependencies of
different statistics by considering the following example. Compare
N-body simulations of four different cosmological models: standard CDM,
open CDM, CDM + $\Lambda$, and  a pure HDM model, the details of which
are given in \S3. Each simulation is normalized to have the same value
of $\sigma_8$, the standard deviation of the mass fluctuations within an
8 $\hmpc$ sphere, where $h$ is the Hubble parameter in units of 100
$\kms~\mpc^{-1}$.  The two point correlation function, $\xi(r)$, for
each model is presented in Figure 1a.  Despite their differences in
initial power spectra, the final correlation functions are similar in
all models.  In general, the correlation function  does not depend
strongly on $\Omega$ if one retains the freedom to set the $\sigma_8$
normalization.  In contrast, Figure 1b shows the pairwise velocity
dispersion, $\avg{v^2}(r)$, as a  function of separation for the same
set of models.  A strong differentiation between the low 
and high $\Omega$ models is apparent.

  The velocity dispersion is easy to compute in simulations, but is
difficult to measure in the geometric projection of phase space that is
measured by redshift surveys. The traditional way to measure the
small-scale velocity dispersion  of galaxies is via the anisotropy it
introduces in the redshift-space two-point correlation function
(\cite{Davis83};  \cite{Fisher94}; \cite{Marzke95}; \cite{Loveday96}). 
However, as it is a pair-weighted statistic, it is dominated by the
densest regions, which  are necessarily rare. Thus one finds large
variance between  estimates of the small-scale velocity dispersion
between different  samples (\cite{Mo93}; \cite{Zurek94}; \cite{Guzzo95};
\cite{Somerville97}), indicating  that the small-scale velocity
dispersion measured in this way is  not very robust.  Attempts have been
made to use direct measures of peculiar  velocities of galaxies as a
constraint on $\avg{v^2}$ (e.g.,  \cite{Strauss93}, \cite{Willick97}),
but other than the nearby universe, where the  velocity field is
observed to be very quiet (\cite{Sandage86};  \cite{Brown87};
\cite{Burstein90}), the errors on the individual  peculiar velocity
measurements swamp the signal from $\avg{v^2}$.

  The primary goal of this paper is to present a statistic---the
redshift dispersion ($\sz$)---that captures $\vsqavg$ in a way which is
naturally applied to volume limited samples drawn from redshift surveys.
In the subsequent sections we motivate and present our redshift
dispersion statistic. Again, because it is a pair-weighted statistic,
computing $\vsqavg$ by averaging over all densities results in a value
heavily weighted by the  densest regions.  However, this problem can be
alleviated if the  dispersion can be calculated {\it as a function of
density}.  The theoretical motivation for our statistic stems from the
Cosmic  Virial Theorem (CVT), derived in Peebles (1976a, hereafter P76),
which relates $\vsqavg$ to $\Om$ and $\xi$.  In \S2 we show that  the
CVT can be applied to subsets of particles in a system which  correspond
to surfaces of constant density.  Many assumptions are necessary to
obtain the results shown in \S2, not all of which are obvious. In \S3 we
explore the relationships derived in \S2 with simple N-body simulations,
suggesting that the results of \S2 hold over a wide range of scales. The
redshift dispersion is entirely independent of the results of \S2 and
\S3, which provide a context for the redshift dispersion that would
otherwise be an unmotivated simulation based statistic. \S4 describes
how to compute the redshift dispersion from a redshift survey. \S5
contains  our conclusions and remarks on future work.

%%%%%%%%%%%%%%%%%%%%%%%%%%%%%%%%%%%%%%%%%%%%%%%%%%%%%%%%%%%%
\section{Velocity Dispersion on Surfaces of Constant Density}
%%%%%%%%%%%%%%%%%%%%%%%%%%%%%%%%%%%%%%%%%%%%%%%%%%%%%%%%%%%%

  As will be shown in \S4, the redshift dispersion probes the pairwise
peculiar velocity dispersion in regions of different density and is
designed for comparing simulated and observed redshift surveys.  In this
section we attempt to provide a theoretical context and some motivation.

  One of the main challenges of working with the pairwise velocity
dispersion arises from its strong density dependence. Intuitively, both
the number of galaxies and the velocity dispersion should be highest in
the densest regions.  Thus, averaging over all densities will give
results dominated by rare, high density peaks.  This problem can be
eliminated if the dispersion is calculated as a function of density.

  In the case of clustered, gravitationally interacting  particles there
exists a scaling relation between the velocity  dispersion on small
scales $\vsqavg (r)$ averaged over pairs  separated by a distance $r$,
and the two point correlation  function $\xi(r)$, the excess fractional
probability of finding  two particles with separation $r$.  This result
is contained in  the CVT derived in P76 (see also \cite{Peebles76b};
\cite{Davis77}; \cite{Peebles80}), which can be written
  \BEQ
        \vsqavg(r) \propto \Om \xi(r) r^2 .
  \EEQ
Several assumptions are used to obtain Eq.~(1) (see P76),  including
that $\xi$ is given by a power law  $\xi(r) = (r_0/r)^\gamma$; $r \ll
r_0$, implying $\xi(r) \gg 1$; the three-point correlation function
$\zeta$ is given by a symmetrized product of the two-point correlation
function (see Eq.~[A23]); and the mass of each galaxy is concentrated on
scales smaller than their separation. The validity of these assumptions
is not obvious (\cite{Fisher94}). For an excellent discussion of the
implications of extended dark matter halos on the CVT, see Bartlett \&
Blanchard (1996).  In addition, we are assuming that the {\it galaxy\/}
velocity field is unbiased with respect to that of the dark matter.
While theoretical investigations have given strong suggestions that a
velocity bias of order 20--30\% may exist (\cite{Couchman92},
\cite{Evrard94}, \cite{Gelb94}), the difficulty in reliably tracing
galaxies in simulations has prevented a good estimate of its magnitude
(\cite{Summers95}).  For the motivational purposes of this paper,
let us keep these assumptions as we work to derive the density
dependence of the CVT.

  To rewrite Eq.~(1) in terms of the density requires some new 
notation.  Let $\xi_{a \times b}(r)$ denote the cross-correlation of 
particle sets $a$ and $b$. The two-point (auto) correlation  function
of all particles in a system $\sP$ can be written  $\xi_{\PcP}(r)$.   The
correlation function can also be written as  the average of the
individual cross correlations
  \BEQ
        \xi(r) = {1 \over N_{\sP}} \sum_{i \in \sP} \xi_{\icP}(r),
  \EEQ
where $N_{\sP}$ is the total number of particles.  We can write  down a
similar expression for the velocity dispersion
  \BEQ
        \vsqavg(r) = {1 \over N_{\sP}} \sum_{i \in \sP} \vicPsqavg (r),
  \EEQ
where $\vicPsqavg (r)$ is the variance of the pairwise velocity  between
particle $i$ and all particles in $\sP$ which lie a distance $r$ from
$i$. We show in Appendix A that these  expressions lead to a
generalization of the CVT that holds for  each particle:
  \BEQ
        \vicPsqavg(r) \propto \Om \xi_{\icP}(r) r^2.
  \EEQ
If the CVT holds for each particle, then it holds for any subset,  $\sS
\subset \sP$: 
  \BEQ
        \vScPsqavg(r) \propto \Om \xi_{\ScP}(r) r^2,
  \EEQ
where
  \BEQ
        \vScPsqavg(r) \equiv {1 \over N_\sS}
          \sum_{i \in \sS \subset \sP} \vicPsqavg(r),
  \EEQ
and
  \BEQ
        \xi_{\ScP}(r) \equiv {1 \over N_\sS}
          \sum_{i \in \sS \subset \sP} \xi_{\icP}(r).
  \EEQ
  Let $N_i(r)$ be the number of particles within a radius $r$ of the $i$th
particle, and $\sS_n$ the set of all particles for which $N_i = n$. 
For this subset, the CVT is
  \BEQ
        \vsqavg(n,r) = C_1 \Om \xi(n,r) r^2 ,
  \EEQ
where $\vsqavg(n,r) \equiv \vSncPsqavg(r)$, $\xi(n,r) \equiv 
\xi_{\SncP}(r)$ and $C_1$ is given by Eq.~(A32).  The quantity
$\xi(n,r)$ is related to the expected number of particles within a
radius $r$ around any particle by
  \BEQ
        \nub \int_0^r [1 + \xi(n,r')] 4 \pi r'^2 dr' = n,
  \EEQ
where $\nub$ is the mean number of particles per unit volume.   If
$\xi(n,r) \propto r^{-\gamma}$ and $\xi(n,r) \gg 1$ (as assumed in  the
original P76 derivation), the above expression yields
  \BEQ
        {3 \over {3 - \gamma}} \xi(n,r) \nb(r) = n,
  \EEQ
where $\nb(r) \equiv \nub \frac{4}{3} \pi r^3$.

  Inserting the above expression into Eq.~(8) yields
  \BEQ
        \vsqavg(n,r) = C_1 C_2 \Om n/\nb(r)^{1/3},
  \EEQ
where $C_2 = (3-\gamma)/3(4\pi\nub/3)^{2/3}$. The above expression
shows that the pairwise velocity dispersion is proportional to $\Om$ and
the local density (through $n$) smoothed on a scale  $r$.  Finally,
since we are working with $N_i(r)$, which is the  number of particles
{\it within} a volume of radius $r$ centered  on a particle, it is
convenient to work with a similar velocity  dispersion.  Let $\svsq(r)$
be the average pairwise velocity  dispersion of all the particles {\it
within} a volume of radius $r$.  $\svsq(r)$ can be obtained by
integrating $\vsqavg(r)$ under  the same assumptions used to derive
Eq.~(1)
  \BEQ
        \svsq(r) \equiv \frac{
          \int_0^r \vsqavg(r') [1 + \xi(r')] 4 \pi r'^2 dr'}
          {\int_0^r [1 + \xi(r')] 4 \pi r'^2 dr'} 
        = C_1 C_3 \frac{\Om \xi(r)^2 r^5}{r^3 \xi(r)}
        = C_3 \vsqavg(r),
  \EEQ
where $C_3 = (3-\gamma)/(5-2\gamma)$. Substituting Eq.~(12) into
(11) results in
  \BEQ
        \svsq(n,r) \propto \Om n/\nb(r)^{1/3}
  \EEQ
where the constant of proportionality is equal to
\BEQ
     C_1 C_2 C_3
         = \frac{3 \Om Q M_\gamma
           (3-\gamma)^2 }
           {4(\gamma - 1)(2 - \gamma)(4 - \gamma)
            (5-2\gamma)(4\pi\nub/3)^{2/3}},
\EEQ
and $Q$ and $M_\gamma$ are defined in Appendix A.

  The above derivation of Eq.~(13) holds whether or not the quantities
are averaged over one or many particles with the same density.  The
above velocity dispersion is computed with respect to the velocity of
the central particle, $\vb$. The velocity dispersion can also be
computed with respect to the mean velocity of the particles in a cell,
$\ub$. In a given cell, these two values of the velocity dispersion will
differ by $|\ub|^2 - |\vb|^2$.  The distribution of these differences
for all cells in $\sS_n$ will peak at zero and have a width on the order
of $\svsq$.  Subsequent averaging over many cells with the same density
will result in zero net difference. Thus, Eq.~(13) also applies for
velocity dispersions computed with respect to the mean velocity in the
cell if the results are averaged over many cells in $\sS_n$.

  Eq.~(13) conveniently relates two readily computable
quantities: the velocity dispersion with respect to the mean velocity in
a cell of radius $r$ to the number of particles in the cell, which is
proportional to the density.  In the next section, we explore the above
form of the CVT with N-body simulations.

%%%%%%%%%%%%%%%%%%%%%%%%%%%%%%%%%%%%%%%%%%%%%%%%%%%%%%%%%%%%
\section{N-Body Results}
%%%%%%%%%%%%%%%%%%%%%%%%%%%%%%%%%%%%%%%%%%%%%%%%%%%%%%%%%%%%

  The previous section presented a derivation for a relationship linking
the velocity dispersion to the local density.  In this section we
explore the range over which Eq.~(13) holds using N-body
simulations of specific cosmological models with different initial power
spectra.  We are particularly interested in the dependence on $\Om$ when
the number of objects is of the same order as expected from volume
limited redshift surveys.

  The simulations we consider are designed to probe a variety of popular
cosmological models.  The four models are: standard CDM  with $\Omega =
1$, $h=0.5$; HDM with $\Omega = 1$, $h=1.0$; open  CDM with $\Omega =
0.35$, $h=0.7$; and CDM + $\Lambda$ with  $\Omega_{CDM} = 0.35$,
$\Omega_{\Lambda} = 0.65$, $h=0.7$.  The open CDM and CDM + $\Lambda$
models provide two alternatives to standard CDM that increase the ratio
of large scale to small scale  power. They differ in evolution in that
structure formation  ``freezes out'' at an earlier epoch in the open
model as the expansion rate exceeds the rate of gravitational collapse.
Thus, to achieve the same level of structure today, collapse must  begin
earliest in the open CDM model, later in the CDM + $\Lambda$  model, and
latest in standard CDM model.  While HDM is not  generally considered a
viable theory, it provides a significantly  different power spectrum
shape with which to explore our ideas.

  The initial conditions are designed to treat the models, as much  as
possible, on an even footing. All models assume a Harrison-Zel'dovich
primordial power spectrum, and use the same random phases for the
Fourier modes to generate the initial density field from their
respective power spectra.  The CDM transfer functions are  taken from
Efstathiou, Bond, \& White (1992, Eq.~[7]) with the parameter
$\Gamma\equiv\Omega_{CDM}\,h$.  Although this function was not intended
for use in open models, it fits more detailed calculations to within 5\%
(D. N. Spergel 1995, private communication).  The HDM transfer function is
taken from Holtzman (1989, Table 2A, line 52).  As stated in \S1, each
model is normalized to have the  same linear value of $\sigma_8=0.67$,
so as to provide similar  correlation strengths and to isolate out the
velocity dependencies.   Although this normalization does not match that
predicted from the  observed fluctuations in the Cosmic Microwave
Background for some  or all of the models (c.f., \cite{Stompor95};
\cite{Gorski95}), it roughly  equalizes the amount of power on the
scales where this paper is focused.

  Each of the simulations follows $32^3=32,768$ dark matter particles
within a periodic cube of comoving size 100 $\hmpc$ (10,000 $\kms$) on a
side. The P3MG3A code (\cite{Brieu95}), which implements the  P$^3$M
algorithm (\cite{Hockney81}; \cite{Efstathiou85}) on the  GRAPE-3A
hardware board (\cite{Okumura93}), is used to evolve the  simulations
from redshift $z=23$ to $z=0$ using 1200 time steps. A  Plummer force
law with softening parameter of 156 $\hkpc$ is used  for the
gravitational interactions and the mass per particle is
$8.5\times 10^{12}~\Omega\,h^{-1}~M_{\sun}$.  Each simulation took
approximately 2 hours to run on a Sun Sparc 10/51 workstation with a 4
chip GRAPE-3A board.

  The ideal way to construct an artificial galaxy catalog is by
identifying concentrations of gaseous and stellar material in high
resolution N-body/hydrodynamic simulations.  Current computer technology
and algorithms now permit such simulations on the scale of small groups
of galaxies (\cite{Evrard94}).  However, at the present time the realm
of large volumes remain the domain of strictly gravitational N-body
codes. Identifying galaxies from dark matter halos is a significant
problem that may not be solvable (\cite{Summers95}); although a variety
of of impressive methods have been developed (\cite{Efstathiou88};
\cite{Bertschinger91}).  However, to keep things as simple as possible
we choose a model in which mass traces light, and each particle is
assumed to be a galaxy.  This approach neglects important processes,
such as mass and velocity bias, the dependence of bias on cosmological
models, and the interaction of dark matter halos. Nevertheless this
approach should be sufficient for our purely motivational purposes. 
Future simulations which can both cover large volumes and resolve
galaxies hydrodynamically will hopefully clarify the nature of the bias.

  To extract the velocity dispersion of a particular surface of constant
density, consider a set of particles where $\xb_i$ and $\vb_i$ are the
position and velocity of the  $i$th particle.  Place spherical cells of
radius $r$ on a uniform grid over the entire domain.  Let $N_j$ be the
number of particles in cell $j$. The correct way to compute the velocity
dispersion in a cell is with respect to the mean motion of the
particles.  As was argued in \S2, Eq.~(13) will apply if the results are
averaged over many cells with the same density. Denote the mean 
velocity in the $j$th cell by $\ub_j$; the velocity variance, 
$\sigma_j^2$, is then $|\vb_i - \ub_j|^2$  averaged over all particles
in the cell. If $\sS_n$ is the set of all cells having $n$ particles 
(\ie $N_j = n$), then the average velocity and variance as a  function
of $n$ is
  \BEQ
        \mu_v(n,r) \equiv {1 \over N_{\sS_n}} \sum_{j \in \sS_n} |\ub_j|.
  \EEQ
and
  \BEQ
        \sigma^2_v(n,r) \equiv
          {1 \over N_{\sS_n}} \sum_{j \in \sS_n} \sigma^2_j,
  \EEQ
where $N_{\sS_n}$ is the number of particles in the set $\sS_n$ (Note:
$N_{\sS_n} \neq n$). These equations provide a specific prescription for
computing the  density dependence of the mean velocity and the velocity 
dispersion, which can readily be applied to the simulations.

  Plots of $\mu_v(n,r)$ and $\sigma_v(n,r)$ are shown in Figure~2 for 
the four models.  The cell size is $r =  194~\kms$, corresponding to
$\nb(r) = 1$. All distances are  expressed in units of $\kms$.  Figure~2
demonstrates three important points: both $\mu_v$ and $\sigma_v^2$  are
independent of the shape of the initial power spectrum;  $\mu_v$ is
independent of the local density, while $\sigma_v^2$ is proportional to
the density; and $\mu_v$ and $\sigma_v^2$ have the  same strong $\Om$
dependence that $\vsqavg$ demonstrated in Figure~1b.

  Formally, the power spectrum independence is explained by the
derivation of the CVT (see Appendix A). The velocity dispersion
depends upon the evolved power spectrum, which is essentially
indistinguishable between models (Figure~1).  Moreover, any remaining
difference between models is encoded in the density distribution
function, which is not apparent when density is the {\it dependent\/}
variable in Figure~2. 

  To explore the range over which Eq.~(13) is valid, $\mu_v$ and  $\svsq$
were calculated over the range of cell sizes $77~\kms \leq r \leq 
488~\kms$, corresponding to $2^{-4} \leq \nb(r) \leq 2^4$.  We  obtained
the following empirical fit for the mean velocity  magnitude
  \BEQ
        \mu_v \propto \Om^{1/2} /\nb^{\alpha} .
  \EEQ
where $\alpha \sim 0.05$.  As $\mu_v$ already contains the desired 
$\Om$ dependence, it is convenient to represent $\sigma_v^2$ in 
terms of the normalized velocity dispersion $\sigma_v^2/\mu_v^2$, 
with the resulting fit
  \BEQ
        \sigma_v^2/\mu_v^2 \propto n/\nb^{1/3} ,
  \EEQ
which agrees with the theory to within the small factor $\nb^{2\alpha}$.

  The quality of the fits can be observed by plotting $\mu_v$ and 
$\svsq/\mu_v^2$ against the scaled density ($n/\nb^{1/3}$), for  each
value of $r$.  Figure~3a shows the ratio of $\mu_v$ to the  fitted value
computed from Eq.~(17) for the $\Om=1$ and $\Om=0.35$  CDM models.  Each
line of data corresponds to a different value of  $r$, and has been
offset from the next by one unit.  Figure~3b shows  a similar plot for the
normalized velocity dispersion  $\szsq/\mu_v^2$.  The solid lines are
the best fits given by  Eq.~(18); this has not been divided out.  The
larger scatter at  higher densities and smaller scales is due to the
small number of  cells contributing to the calculation at these values. 
The key point to observe from Figure~3a is how well the data points
follow the horizontal lines corresponding to their fitted values.

  The fits to the data shown in Figure~3 are remarkable, 
considering that the range includes cells with  underdensities ($n/\nb -
1$) of $-0.8$ on scales of $488~\kms$ and  cells with overdensities of
nearly 200 on scales of $77~\kms$.   Furthermore, the scaling of
$\sigma_v^2$ is almost exactly that  derived from the CVT, indicating
that Eq.~(13) holds over a large  range of scales and densities, even
when $\xi(r) \sim 1$.

%%%%%%%%%%%%%%%%%%%%%%%%%%%%%%%%%%%%%%%%%%%%%%%%%%%%%%%%%%%%
\section{Redshift Dispersion}
%%%%%%%%%%%%%%%%%%%%%%%%%%%%%%%%%%%%%%%%%%%%%%%%%%%%%%%%%%%%

  The purpose of \S2 was to theoretically motivate the local density and
$\Om$ dependence of $\svsq$.  In \S3 the scaling relations in \S2 were
explored with simple N-body simulations.  In addition, \S\S2 and 3 have
introduced the ideas which allow us to describe the main point of this
paper---the redshift dispersion.

  The formalism we have developed so far cannot be directly applied to 
observations due to the geometric projection of a six dimensional phase
space into a three dimensional redshift survey.  The redshift of
galaxies represents the only probe, albeit indirect, of the peculiar
velocity. Thus, any statistic that desires to take advantage of the
properties of $\svsq$  must be defined with the specific geometry of
redshift space in mind.  In this section we now describe the redshift
dispersion, a statistic with the aim of being readily measurable from
volume limited samples taken from redshift surveys and which captures
the essence of $\svsq$.

  Now let us define quantities analogous to those in \S3, but for
redshift space.  Consider a volume limited sample from a survey out to a
maximum  redshift of $Z$.  Each data point consists of two angular
coordinates on the  celestial sphere and a redshift.  Let us define cells
within which  to measure density and velocity dispersion in projection
on the  celestial sphere. The cell $j$ consists of a cone emanating from
 the origin with solid angle $\pi \theta^2$ around the cell center.  
The number of points in the cone $j$ is $N_j$ and is proportional to the
projected density on the sky.  The mean and the variance of  the
redshifts in the cone are denoted $u_j$ and $\sigma_j^2$, which are
depicted schematically in Figure 4.  If  $\sS_n$ is the set of all cones
with $N_j = n$, then the average  and variance of the redshift as a
function of $n$ is
  \BEQ
       \mu_z(n,\theta) \equiv
         {1 \over N_{\sS_n}} \sum_{j \in \sS_n} u_j.
  \EEQ
and
  \BEQ
       \sigma^2_z(n,\theta) \equiv
         {1 \over N_{\sS_n}} \sum_{j \in \sS_n} \sigma^2_j.
  \EEQ
in analogy with Eqs.~(15) and (16).

  The efficacy with which the $\sz$ statistic might distinguish between
models is examined with the simulations discussed in \S3.  The
simulations were transformed into redshift-space using $Z = 5000~\kms$,
which is equal to one half of the simulation box size.  Since the data
are periodic, the origin can be placed at any point, allowing multiple
perspectives to be drawn from a single simulation.  We computed
$\sigma_z$ for $\Om = 1.0$ and $\Om = 0.35$ CDM models.  The angular
cell size was $\theta =1.8 \arcdeg$  ($\nb(\theta) = 2$), and the cell
centers were computed by  creating a pseudo-uniform grid of points
across the celestial  sphere (Baumgardner \& Frederickson 1985).  The
data have been  averaged over 27 different origins regularly distributed
throughout the domain, and error bars computed from the standard 
deviation of this averaging.

  The $\sz$ curves are plotted in Figure~5.  As with the velocity
dispersion in Figure~1b, the redshift dispersion shows a strong
separation between  the low and high $\Om$ models.  The differences
become apparent at moderate angular over-densities ($\delta\sim 5$). The
shape of the curves in Figure~5 is due to the combined spatial and
peculiar velocity contributions to $\sz$, as is illustrated in Figure~6
for the $\Omega = 1$ CDM simulation.  We know the full six-dimensional
position of each galaxy in phase space in the N-body simulation, and
thus can separate these two contributions.  At low overdensities, $\sz$
is dominated by the spatial separation of the particles, as is indicated
by the solid points in Figure~6; the spatial component scales
approximately as $n^{-1/2}$, due to the more tightly bound nature of
denser systems. At higher overdensities the peculiar velocity dominates,
scaling approximately as $n^{1/2}$, which is consistent with results of
\S2 and \S3.  For smaller values of $\Om$ the spatial component behaves
the same, but the peculiar velocity component is down by a factor of
$\Om^{1/2}$.

  These results indicate that the greatest separation in the dispersion
between models with different values of $\Om$ will occur in the denser
regions; so it is important that we sample many modestly dense regions,
which requires a large volume. In addition, to minimize projection
effects the sample should not be too deep (i.e., $Z$ not  too large).
Therefore, to apply the $\sz$ statistic requires a dense sample over a
wide field. If believable simulations of galaxies can be developed it
might be possible to constrain models with the same final correlation
function by varying $\Omega$ in the simulations and finding the best fit
to the observations.  For any one point, Figure~5 indicates an error of
about 0.15 in $\Omega$.  Using many points along the curve, the errors
may be small enough to significantly constrain the value of $\Omega$.

  The current redshift surveys do not have enough data to attempt such a
comparison.  To get the level of  separation shown in Figure~5 required
${\cal O}(10^4)$ particles.  We have calculated $\sigma_z$ from the IRAS
1.2 Jansky survey (\cite{Fisher95}), but a volume-limited sample taken
from this survey contains only  800 galaxies at best. The error bars
from an 800 point sample in the simulations were too large to be able to
distinguish between high and low $\Om$ models.

  Fortunately, a substantial increase in the amount of data  available
will be brought about by the Sloan Digital Sky Survey  (SDSS).  The SDSS
will obtain spectra and measure the redshifts  of the $10^6$ galaxies
down to $r' \sim 18$  (\cite{Gunn95}). We can estimate  the size of a
volume-limited sample taken from the SDSS using the Schechter luminosity
function $\phi_s$ fitted to the Stromlo-APM survey  (\cite{Loveday92})
  \BEQ
        \phi_s(L) dL = \phi^{\star} y^{\alpha} e^{-y} dy, ~~~
                   y = L/L^{\star},
  \EEQ
where $\phi^{\star} \simeq 1.4\times 10^{-8}\,{\rm km^{-3}\,s^3}$, 
$M^{\star} \simeq -19.5$, and $\alpha \simeq -0.97$ are parameters 
obtained from the fit.  The number of galaxies in a given volume, 
$V$, brighter than $L_0$, $N(V,L_0)$, can be computed by 
integrating the luminosity function
  \BEQ
        N(V,L_0) = V \int_{L_0}^{\infty} \phi(L) dL 
                 = V \int_{y_0}^{\infty} 
                   \phi^{\star} y^{\alpha} e^{-y} dy
                 = V \phi^{\star} \Gamma(\alpha + 1,y_0),
  \EEQ
where $y_0 = L_0/L^{\star}$.  For a volume limited survey, $L_0$  is the
luminosity an object would have if it had an apparent  magnitude $m_0$
and was located at the volume edge $Z$ ($H_0 =  100\,\kms\,\mpc^{-1}$). 
From this definition it follows that
  \BEQ
        y_0 = Z^2_{\kms}\,10^{0.4 (M^{\star} - m_0 +15)},
  \EEQ
where $m_0$ is the apparent magnitude limit of the survey. The
Stromlo-APM survey was taken in the $b_j$ band, while the  Sloan will
use the $r'$ band.  The two can be equated by the  approximate relation
$b_j \sim r' + 1$.  However, we set $m_0 =  18.7$ which gives a better
value for the estimated total number of  galaxies in the survey
($10^6$).  Inserting $Z = 5000\,\kms$ into  Eq.~(23) gives $y_0 \approx
0.013$.  Setting $V_{SDSS} = \pi Z^3/3$  results in $N_{SDSS} \approx
7000$. This number might in fact be appreciably larger if there are 
many more faint galaxies than Eq.~(22) implies, and has been  suggested
by recent surveys for low-surface brightness galaxies  (cf.,
\cite{Dalcanton95}).  However, these galaxies are unlikely  to have
redshifts measured as part of the SDSS.

  One could substantially increase the number of galaxies in a
volume-limited survey of fixed depth by dropping the requirement that it
include the origin. Figure~7 plots the expected number of galaxies in a
series of volume limited shells for SDSS and shows that a redshift shell
between $25,000\,\kms$ and $30,000\,\kms$ would include almost $10^5$
galaxies!

  Ultimately, applying the redshift dispersion statistic to the SDSS
could provide many data points for comparing with simulations and
perhaps constraining $\Omega$.  The $\sz$ curves can be evaluated for
many angular sizes and redshift shells for both the SDSS and for several
next generation, high resolution, N-body/hydrodynamic simulations with
different values of $\Om$.  Since the number of estimated objects in the
SDSS is the same order or more as the simulations shown in Figure~5, one
might expect to obtain an equivalent separation between models for each
value.  It should be interesting to compare $\sz$ measured in different
simulations with the SDSS as a function of $n$, $\theta$ and shell
geometry.

%%%%%%%%%%%%%%%%%%%%%%%%%%%%%%%%%%%%%%%%%%%%%%%%%%%%%%%%%%%%
\section{Conclusions}
%%%%%%%%%%%%%%%%%%%%%%%%%%%%%%%%%%%%%%%%%%%%%%%%%%%%%%%%%%%%

  Our goal has been to present a new statistic---the redshift dispersion
($\sz$)---that is sensitive to $\Om$ and is well suited for comparing
real and simulated volume limited samples from redshift surveys. Given
the proper data, $\sz$ is easy to compute and can be applied on many
scales without applying arbitrary assumptions.  We have used low
resolution simulations, which are sufficient for our motivational
purposes, to do a simple exploration of $\sz$, which suggests that
applying it to the SDSS may be worthwhile. In addition, with the right
simulations, it might be possible to constrain $\Om$ in models where the
simulations match the observed final correlation function.

 We have shown that the pairwise velocity dispersion is  intrinsically
related to the local density and $\Om$. In \S2 and Appendix A it is
shown that the CVT holds for each particle in a system and subsequently
for any subset of  particles, if we are careful to define the velocity
dispersion and correlation function for these subsets appropriately.  
The density dependence of the velocity dispersion can be extracted  by
looking at subsets of particles of a given local density (see Eq.~13).
Knowing how $\svsq$ depends upon the local density gives  an indication
as to why the standard approach of averaging $\vsqavg$ over all
densities is highly sensitive to the presence of rare  density peaks. 
Exploring Eq.~(13) with N-body simulations of  several cosmological
models indicates that it holds over a wide  range of length scales,
$77~\kms \leq r \leq 488~\kms$.  Our redshift dispersion statistic is
simply the redshift-space analog of the quantity $\sv$.

  We have treated our galaxies as equal mass particles containing all
the mass, ignoring the hypothesized global stochastic mapping from the
dark matter mass and velocity distribution to the distribution of
galaxies (i.e. the mass and velocity bias functions), which may differ
among the models.  A more general treatment would estimate the mass and
velocity bias in each model and normalize the models such that the
galaxy correlations, not the dark matter correlations, are similar.  At
best, reliable estimates of these bias functions await the next
generation of simulations where galaxies can be resolved within
statistically meaningful volumes.  However, there are several arguments
as to why the differences in the biases between models may not strongly
affect our work.  First, the initial power spectra of currently favored
hierarchical models all have similar slopes on galaxy formation scales.
Hence, the initial conditions of galaxy formation---and the resulting
bias---may be similar.  In addition, the HDM model, which has a very
different initial power spectrum, has a similar final correlation
function, which suggests that the final power on small scales is
dominated by non-linear processes which erase initial differences. Any
velocity bias which arises through dynamical friction should be similar
for models evolved to similar clustering levels.  If velocity bias is
related to galaxy formation sites preferentially near potential wells,
then velocity bias could depend on mass density and the efficacy of this
measure could be diminished.  However, both types of bias appear to be
strong only in the most non-linear regions ($\delta\gtrsim200$)
(\cite{Summers95}), while our study focuses on the mildly non-linear
regimes ($\delta\sim10$).

  High resolution, hydrodynamic simulations that include detailed galaxy
formation should improve our understanding of bias and how it changes
from model to model. Intuitively, any bias, by eliminating objects in
the underdense regions, should have the overall effect of shifting the
data points in Figure~5 to the left.  Higher resolution will also
incorporate the interactions of dark matter halos, which may
significantly effect the three point correlation function on galaxy
scales (see \cite{Bartlett96}).  All of these effects indicate more
detailed simulations, beyond what is currently available, may be
necessary for the actual application of the redshift dispersion.

  The next logical step is to attempt to apply $\sz$ to denser redshift
surveys than the IRAS 1.2 Jansky survey. In the mean time, additional
studies on larger volume, higher resolution N-body simulations would be
useful, but perhaps overkill until a suitable redshift survey becomes
available.  Also, it is not clear that using dark matter halos without
the proper means for identifying galaxies would add to these results. 
Further exploration should also be done on a wide variety of survey
geometries. Here, we only looked at a single small sphere.  It is quite
possible that a larger sphere or a shell might be the  optimal shape to
balance the tradeoff between high density and  large numbers of clusters
that make $\sz$ work best.

%%%%%%%%%%%%%%%%%%%%%%%%%%%%%%%%%%%%%%%%%%%%%%%%%%%%%%%%%%%%
\acknowledgments We would like to thank David Spergel and our
editor Ed Bertschinger for their helpful comments, and John
Baumgardner for the use of his icosahedral mesh code. We gratefully
acknowledge the Grand Challenge Cosmology Consortium and grants of
computer time at the San Diego and Pittsburgh  Supercomputer Centers. 
MAS acknowledges the support of the WM  Keck Foundation, the Alfred
P. Sloan Foundation, and NASA
Astrophysical Theory Grant NAG5-2882. This work was
supported by NSF grants ASC 93--18185 and AST 91--08103.

\appendix

%%%%%%%%%%%%%%%%%%%%%%%%%%%%%%%%%%%%%%%%%%%%%%%%%%%%%%%%%%%%
\section{Cosmic Virial Theorem}
%%%%%%%%%%%%%%%%%%%%%%%%%%%%%%%%%%%%%%%%%%%%%%%%%%%%%%%%%%%%

In \S2 we showed that the CVT applied to sets of particles  provided it
holds for individual particles.  To prove the latter result  we return
to Peebles' original derivation (see P76).  \cite{Davis77} later
rederived the Cosmic Virial Theorem by placing it in  the context of the
BBGKY hierarchy.  We choose to  use the earlier approach because of its
simpler, more intuitive  nature.   The essential derivation is still
that found in P76, but a   few minor modifications have been added to
elaborate key steps in  the derivation.

  The derivation of the CVT proceeds in the following manner.   First, an
equation linking the velocity, acceleration, and  correlation function
of a general particle system is obtained from  the conservation of phase
space density.  Second, the  accelerations are linked through gravity to
the two and three  point correlation functions.  Third, the assumptions
of a  cosmological system are applied to the general particle equation 
to obtain the scaling between the velocity dispersion and the  density. 
The key modification that we have made in order to show  that this
result applies for individual particles in the system is  to replace the
entire phase space with the phase space of a single  particle.

%%%%%%%%%%%%%%%%%%%%%%%%%%%%%%%%%%%%%%%%%%%%%%%%%%%%%%%%%%%%
\subsection{Phase Space Conservation}
%%%%%%%%%%%%%%%%%%%%%%%%%%%%%%%%%%%%%%%%%%%%%%%%%%%%%%%%%%%%

To start, consider a system of particles with the position,  velocity,
and acceleration of the $i$th particle given by $\xb_i$,  $\vb_i$, and
$\ab_i$.  Strictly speaking, these are not phase space variables, but
represent a three-dimensional single-particle distribution and its
evolution in time.  Now consider any two particles $i$ and $j$  having
relative position, velocity, and acceleration:  $\rb = \xb_j - \xb_i$,
$\vb = \vb_j - \vb_i$, $\ab = \ab_j -  \ab_i$, as shown in Figure~8a. 
The rate of change in the distance,  $r$, between the two particles
obeys
  \BEQ
        \rd = v_r = \rdotv / r,
  \EEQ
and the acceleration
  \BEQ
        \rdd = (\rdota + v^2 - (\rdotv/r)^2)/r = (\rdota + v_t^2)/r,
  \EEQ
where $v^2 = v_r^2 + v_t^2$.  For a sufficiently large system, the 
number of particles found in a volume of phase space centered on  the
$i$th particle is $dN_{\icP} = f_{\icP}(r,\rd,\rdd) dr d\rd  d\rdd$,
where $f_{\icP}$ is the phase space density of all pairs  connected to
the particle $i$.  The function $f_{\icP}$ is related  to the two point
correlation function by
  \BEQ
        dr \int d\rd d\rdd f_{\icP} = \nub (1 + \xi_{\icP}) 4 \pi r^2 dr,
  \EEQ
where $\nub$ is the mean number density and $\xi_{\icP} =  \xi_{\icP}(r)$
is the isotropic two point correlation function of  the $i$th particle,
which is defined with respect to the joint  probability $dP_2 = dV_i
dV_j \nub^2 (1 + \xi_{\icP})$.  

For convenience, we now explicitly drop the subscript $\icP$  notation:
$f_{\icP} \rightarrow f$, $N_{\icP} \rightarrow N$, and  $\xi_{\icP}
\rightarrow \xi$.  It should be understood that all  these quantities
are relative to the $i$th particle, including the  velocities and
accelerations.

Having defined the particle system and the phase space density,  $f$, it
is now possible to look at the time evolution of the  particle within
the system.  The integral of $f$ links it to $\xi$  giving the total
number of particles with separation between $r$  and $r+\dr$.  The
number of particles with separation less than  $r$ is
  \BEQ
        N(r,t) = \int_0^r dr \int d\rd d\rdd f.
  \EEQ
At some later time $\dt$ there exists a $\dr$ such that
  \BEQ
        N(r,t) = N(r + \dr,t + \dt).
  \EEQ
Without loss of generality, we could have just as well started at  the
point $N(r - \dr,t)$, and chosen $\dt$ and $\dr$ to satisfy
  \BEQ
        N(r - \dr,t) = N(r,t + \dt),
  \EEQ
which allows us to expand around $r$ and $t$.  Subtracting  $N(r,t)$
from the above expression, we have
  \BEQ
        -[N(r,t) - N(r - \dr,t)] = N(r,t + \dt) - N(r,t).
  \EEQ
The right side can be readily expanded in powers of $\dt$ and is  simply
$\dt \Nd + \half \dt^2 \Ndd$.  Likewise, from the  definition of
$N(r,t)$ the left side is given by
  \BEQ
        -[N(r,t) - N(r - \dr,t)] = - \int d\rd d\rdd \int_{r-\dr}^{r} dr f.
  \EEQ
The Taylor expansion of the integral gives
  \BEQA
        \int_r^{r'} dr f & \simeq & \half \dr [f(r) + f(r-\dr)] \NN \\
                         & \simeq & \dr f - \half \dr^2 \pr f \NN \\
                         & \simeq & \dt \rd f + \half \dt^2 (\rdd f - \rd^2 \pr f),
  \EEQA
where the last step made use of the following expansion for $\dr$
  \BEQ
        \dr = \dt \rd + \half \dt^2 \rdd + \ldots,
  \EEQ
which is valid for {\it fixed} $\rd$ and $\rdd$.

Matching powers of $\dt$ gives an expression for $\Ndd$
  \BEQA
         \Ndd & = & - \int d\rd d\rdd (\rdd - \rd^2 \pr) f \NN \\
              & = & - \int d\rd d\rdd\, \rdd f  + \pr \int d\rd d\rdd\, \rd^2 f 
                   \NN \\
              & = & -  \nub 4 \pi [r^2 \xo \rddavg - \pr (r^2 \xo \rdsqavg)],
  \EEQA
where we have used $\pr \rd = 0$, but $\pr \rdsqavg \neq 0$.  The  time
derivative of $N$ can be written
  \BEQ
        \Ndd  =  4 \pi  \nub \pdd \int_0^r dr \,r^2 \xo
              =  4 \pi  \nub \int_0^r dr \,r^2 \xidd,
  \EEQ
which is combined with the previous expression to give
  \BEQ
        \Ndd \propto \pr [r^2 \xo \rdsqavg] - r^2 \xo \rddavg
              =  \int_0^r dr\, r^2 \xidd.
  \EEQ
Substituting in the expressions for $\rd$ and $\rdd$, and  expanding the
$r$ derivative results in the general equation  derived in P76
  \BEQ
        r \pr [\xo \vrsqavg] + \xo \vrvtsqavg =
           \xo \rdotaavg + r^{-1} \int_0^r dr\, r^2 \xidd.
  \EEQ
This equation represents the conservation of phase space for a  single
particle in a system, and is true for any particle in the  system with a
well defined phase space density and an isotropic  two-point correlation
function.

%%%%%%%%%%%%%%%%%%%%%%%%%%%%%%%%%%%%%%%%%%%%%%%%%%%%%%%%%%%%
\subsection{Gravitating Particles}
%%%%%%%%%%%%%%%%%%%%%%%%%%%%%%%%%%%%%%%%%%%%%%%%%%%%%%%%%%%%

For gravitationally interacting particles, the acceleration term  is
simply $\rdotaavg = \rdotgavg$, where
  \BEQ
        \gb_i = G \sum_j m_j \frac{\xb_j - \xb_i}{|\xb_j - \xb_i|^3}.
  \EEQ
When averaging over phase space, $\gb_i$ can be written in terms  of the
two and three point correlation functions, which requires  considering a
third particle at $\xb_k$ (see Figure~8b).  The  average force on $i$ is
the force from $j$ plus the force due to  $k$ weighted by the
conditional probability of the third particle  being located at $\sb$
  \BEQA
        \gbiavg & = & \frac{Gm \rb}{r^3}
                      + Gm \int \frac{\sb}{s^3} \frac{dP_3}{dP_2}
                      \NN \\
                & = & \frac{G m \rb}{r^3} + \frac{G \rho}{1+\xi}
                      \int d^3 \sb \frac{\sb}{s^3}
                      [1 + \xi(r) +\xi(s) + \xi(q) + \zeta(r,s,q)]
  \EEQA
where $m$ is the average mass per particle and the three point 
correlation function $\zeta$ is defined by the probability
  \BEQ
        dP_3 = dV_i dV_j dV_k \nub^3
               [1 + \xi(r) + \xi(s) + \xi(q) + \zeta(r,s,q)].
  \EEQ
The first, second and third terms in the brackets integrate to  zero by
isotropy leaving
  \BEQ
        \gbiavg =   \frac{Gm \rb}{r^3} + \frac{G \rho}{1+\xi}
                    \int d^3 \sb \frac{\sb}{s^3} [\xi(q) + \zeta(r,s,q)] .
  \EEQ
Likewise, the force on $j$ is
  \BEQ
        \gbjavg = - \frac{Gm \rb}{r^3}
                  - \frac{G \rho}{1+\xi}
                  \int d^3 \qb \frac{\qb}{q^3} [\xi(s) + \zeta(r,q,s)].
  \EEQ
Because $\zeta(r,s,q) = \zeta(r,q,s)$, the average  of
the total force is
  \BEQ
        \rdotgavg = - \frac{2Gm}{r} - \frac{2G \rho}{1+\xi}
                    \int d^3 \sb \frac{\rdots}{s^3} [\xi(q) + \zeta(r,s,q)].
  \EEQ

%%%%%%%%%%%%%%%%%%%%%%%%%%%%%%%%%%%%%%%%%%%%%%%%%%%%%%%%%%%%
\subsection{Cosmological Scalings}
%%%%%%%%%%%%%%%%%%%%%%%%%%%%%%%%%%%%%%%%%%%%%%%%%%%%%%%%%%%%

To obtain the scaling of the velocity dispersion we apply various 
approximations that are consistent with a cosmological model.   First,
we consider the situation where the correlation function is  assumed to
be stationary in comparison to the motions of the  particles, \ie the
time scale for changes in $\xi$ is much greater  than the crossing time
$r/\vrsqavg^{1/2} \rightarrow \xidd = 0$.   In addition, $\vb$ is
randomly oriented so that $\vsqavg/3 \simeq  \vtsqavg/2 \simeq
\vrsqavg$, and $\vrvtsqavg = 0$.  These  assumptions eliminate two terms
from Eq.~(A14).  Combining with Eq.~(A20) gives
  \BEQ
        r \pr [\xo \vsqavg] = -\frac{6Gm \xo}{r} - 6 G \rho
                              \int d^3 \sb \frac{\rdots}{s^3}
                              [\xi(q) + \zeta(r,s,q)].
  \EEQ
The next step is to evaluate the integrals on the right.  At this  point
we impose a model for $\xi$ and $\zeta$.  Observations  indicate that
the correlation functions are well modeled by power  laws of the form
  \BEQ
        \xi(r) = \left(\frac{r_0}{r}\right)^\gamma, ~~~ 
        \gamma = 1.77, ~~~ 
           r_0 = 5.4 \hmpc
  \EEQ
and,
  \BEQ
        \zeta(r,s,q) = Q[\xi(r) \xi(s) + \xi(r) \xi(q) + \xi(s) \xi(q)], ~~~
        Q \simeq 1.0.
  \EEQ
Using these relationships, the first integral can be solved with  the
following identities $\rdots = r s \cos \theta$,  and $q^2 =  r^2 - 2 s
r \cos \theta + s^2$, and by noting that in spherical  coordinates $d^3
\sb = s^2 ds \sin \theta d\theta d\phi$
  \BEQA
        \int d^3 \sb \frac{\rdots}{s^3} \xi(q)
             & = & \int s^2 ds \sin \theta\, d\theta d\phi 
                   \frac{r s \cos \theta}{s^3} \xi(q)
                   \NN \\
             & = & 2 \pi \int ds\, d\theta\, r \cos \theta \sin \theta\,
                   \xi( (r^2 - 2 s r \cos \theta + s^2)^{1/2})
                   \NN \\
             & = & 2 \pi r^2 \int dy\, d\theta \cos \theta \sin \theta\,
                   \xi( (r^2 - 2 y r^2 \cos \theta + y^2 r^2)^{1/2})
                   \NN \\
             & = & 2 \pi \xi r^2 \int_0^{\infty} dy \int_0^{\pi}
                   \frac{d\theta \cos \theta \sin \theta}
                   {(1 - 2y \cos \theta + y^2)^{\gamma/2}} 
                   \NN \\
             & = & \frac{2 \pi \xi r^2 J_{\gamma}}{(2 - \gamma)(4 -\gamma)}.
  \EEQA
where $J_{\gamma} = \int_0^{\infty} dy I_{\gamma}(y)/y^2$, $y =  s/r$,
and
  \BEQ
        I_{\gamma}(y) = (y+1)^{2 - \gamma} \{1 - (2-\gamma) y + y^2\} - 
                        |y-1|^{2 - \gamma} \{1 + (2-\gamma) y + y^2\}
  \EEQ
For the second integral, we obtain a similar result
  \BEQA
        \int d^3 \sb \frac{\rdots}{s^3} \zeta(r,s,q)
             & = & \int d^3 \sb \frac{\rdots}{s^3}
                   Q [\xi(r) \xi(s) + \xi(r) \xi(q) + \xi(s) \xi(q)]
                   \NN \\
             & = & Q \int d^3 \sb \frac{\rdots}{s^3} \xi(q) [\xi(r)  + \xi(s)] 
                   \NN \\
             & = & \frac{2 \pi Q \xi^2 r^2 M_\gamma}{(2 - \gamma)(4 - \gamma)},
  \EEQA
where $M_\gamma = \int_0^{\infty} dy (1 + y^{-\gamma}) 
I_{\gamma}(y)/y^2$.  For our purposes, it is sufficient to know  that
$J_{1.8}$ and $M_{1.8}$ are of order unity.  Inserting the  above
results into Eq.~(A22) gives
  \BEQ
        r \pr [\xo \vsqavg] = - \frac{6Gm \xo}{r}
                              - \frac{12 \pi G \rho\xi r^2 J_{\gamma}}
                                {(2 - \gamma)(4 -\gamma)}
                              - \frac{12 \pi G \rho Q \xi^2 r^2 M_\gamma}
                                {(2 - \gamma)(4 - \gamma)}.
  \EEQ
Finally, let $\xi \gg 1$.  Dividing through by $r$, and  integrating
from $r$ to $\infty$ and dividing again by $\xi$,  gives an equation
for the velocity dispersion
  \BEQ
       \vsqavg(r) = \frac{6Gm}{(\gamma + 1) r}
                    + \frac{12 \pi G \rho r^2 J_{\gamma}}
                      {(2 - \gamma)^2 (4 -\gamma)}
                    + \frac{6 \pi G \rho Q \xi(r) r^2 M_\gamma}
                      {(\gamma - 1)(2 - \gamma)(4 - \gamma)}
  \EEQ
For small $r$ the third term dominates the second.  The first term is
negligible in the limit that $m$ is very small, i.e., that the mass is
divided up into many tiny individual particles.  Thus 
%ratio of the
%first term to the third results in an expression that is $\sim m/\rho
%r^3 \xi$. This ratio will be small when there are many particles within
%$r$ for a pair at separation $r$, which if we assume to be true
%indicates that 
the third term dominates, leaving the classic result
  \BEQ
       \vsqavg(r) = \frac{6 \pi G \rho Q \xi(r) r^2 M_\gamma}
                    {(\gamma - 1)(2 - \gamma)(4 - \gamma)}.
  \EEQ
Using the definition $H^2 = 8\pi G \rho/ 3 \Om$, we can rewrite  the
above expression as
  \BEQ
        \vsqavg(r) = \frac{9 \Om Q (Hr)^2 \xi(r) M_\gamma}
                     {4(\gamma - 1)(2 - \gamma)(4 - \gamma)}.
  \EEQ
which, if we return to the earlier notation (measuring distances in
terms of velocity $H r$) gives the expression  quoted in Eq.~(4) of \S2
  \BEQ
        \vicPsqavg(r) \propto \Om \xi_{\icP}(r) r^2,
  \EEQ
with a proportionality constant
  \BEQ
        C_1 = \frac{9 \Om Q M_\gamma}
                   {4(\gamma - 1)(2 - \gamma)(4 - \gamma)}.
  \EEQ

\newpage

%%%%%%%%%%%%%%%%%%%%%%%%%%%%%%%%%%%%%%%%%%%%%%%%%%%%%%%%%%%%

\newpage

%%%%%%%%%%%%%%%%%%%%%%%%%%%%%%%%%%%%%%%%%%%%%%%%%%%%%%%%%%%%

\begin{figure}
\plotfiddle{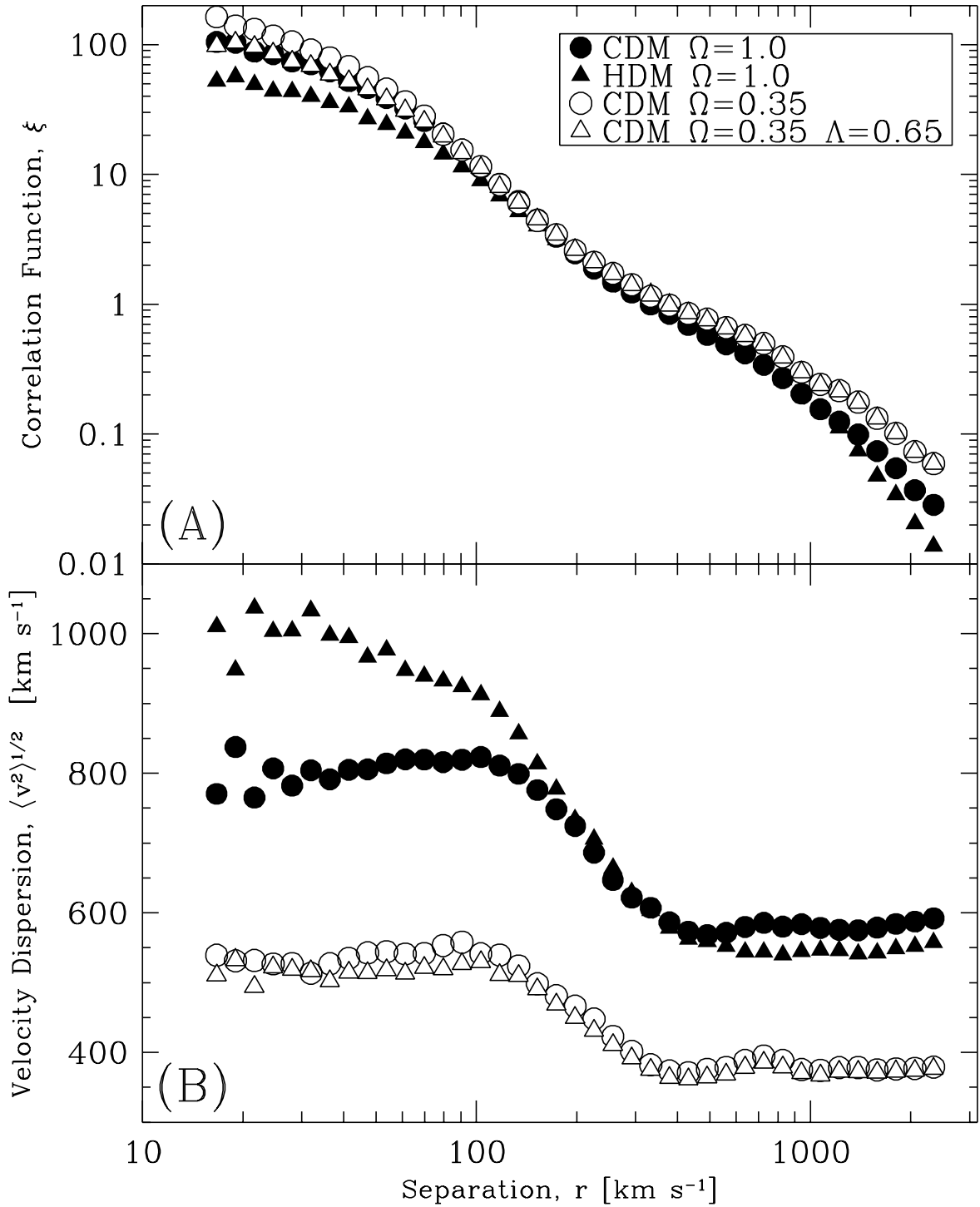}{19cm}{0}{90}{90}{-300}{-100}
%\plotone{fig1.eps}
\caption{
  Two-point correlation function {\bf (a)} and pairwise  velocity
dispersion {\bf (b)} measured in N-body simulations of  four
cosmological models.  Each model was normalized to have  similar power
on small scales (\ie $\sigma_8 = 0.67$).
}
\end{figure}

\begin{figure}
\plotfiddle{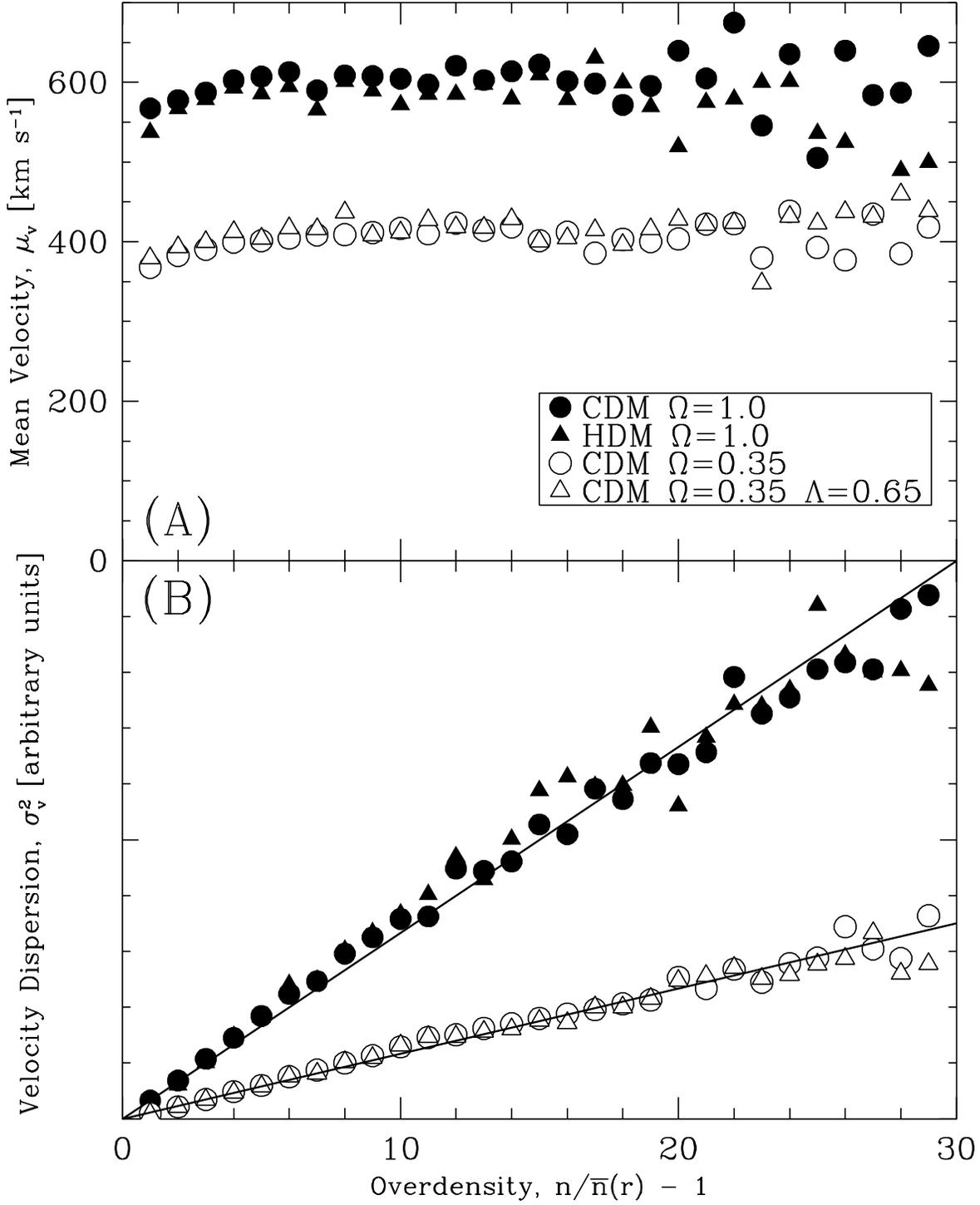}{19cm}{0}{90}{90}{-300}{-100}
\caption{
  {\bf (a)} Mean flow $\mu_v$ and {\bf (b)} velocity  variance $\svsq$
as a function of overdensity, $n/\nb - 1$, for N-body models.  These
data were computed using $10^5$ cells with $r  = 194\, \kms$ and $\nb =
1$.
}
\end{figure}

\begin{figure}
\plotfiddle{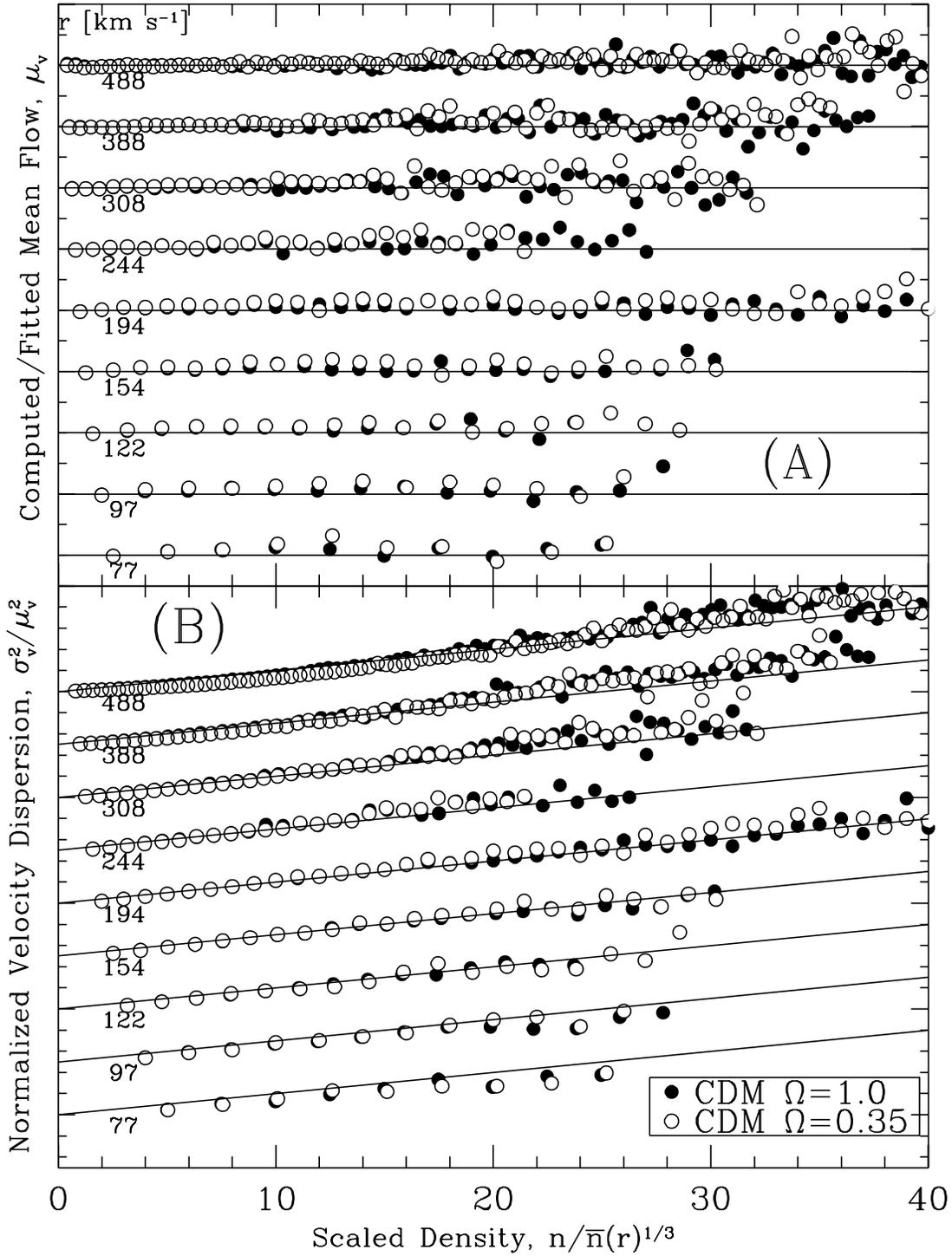}{19cm}{0}{90}{90}{-300}{-100}
\caption{
  Empirical fits to {\bf (a)} $\mu_v$ and {\bf (b)}  $\svsq$, as given
by Eqs.~(17) and (18).  The vertical axis has  been scaled so that each
line of points corresponds to a different  cell size.  From top to
bottom the cell sizes are $r (\kms) =  [488, 388, 308, 244, 194, 154,
122, 97, 77]$ corresponding to  $\nb(r) = [2^4, 2^3, 2^2, 2^1, 2^0,
2^{-1}, 2^{-2}, 2^{-3}, 2^{- 4}]$.  The horizontal axis is in terms of
scaled units  $n/\nb^{1/3}$.  The lines indicate the value obtained from
the  empirical fits.
}
\end{figure}

\begin{figure}
\plotfiddle{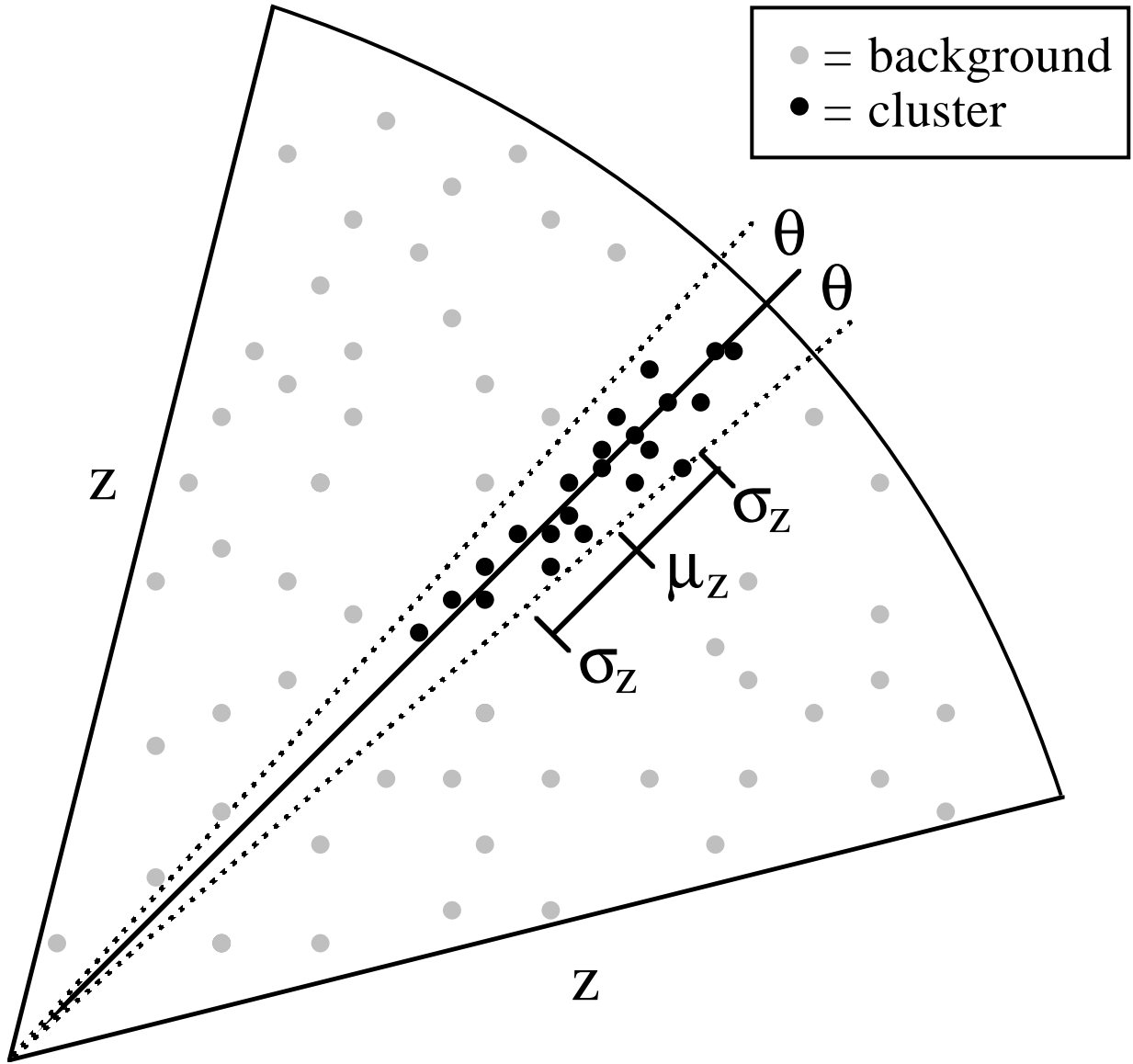}{13cm}{0}{150}{150}{-325}{-600}
\caption{
  Schematic illustration of the mean redshift, $\mu_z$,  and the
redshift dispersion, $\sz$, in a pie slice projection of a  cluster.
}
\end{figure}

\begin{figure}
\plotfiddle{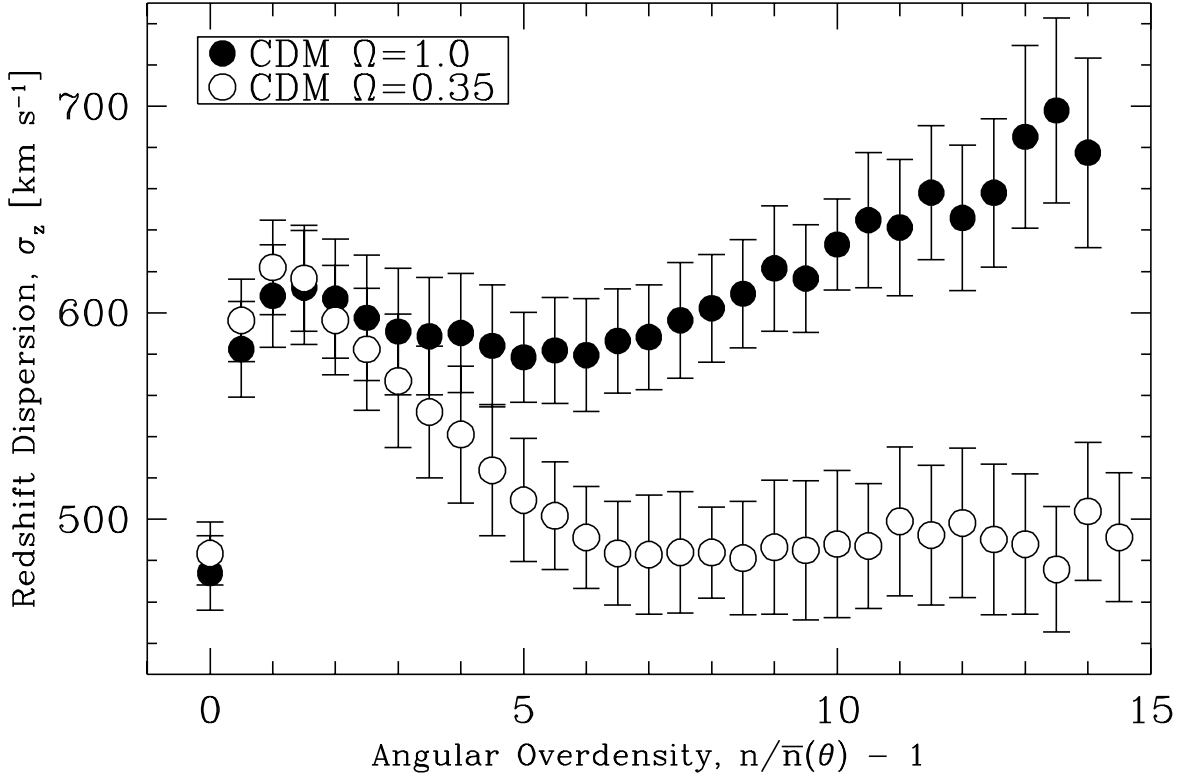}{10cm}{0}{90}{90}{-300}{-350}
\caption{
  Redshift dispersion, $\sz$, as a function of angular  overdensity for
8000 dark matter particles out to $Z = 5000~\kms$  averaged over 27
viewpoints.  The error bars are the standard  deviation of these
averages.  The angular radius of the cell was  $\theta = 1.8 \arcdeg$,
corresponding to $\nb = 2$.
}
\end{figure}

\begin{figure}
\plotfiddle{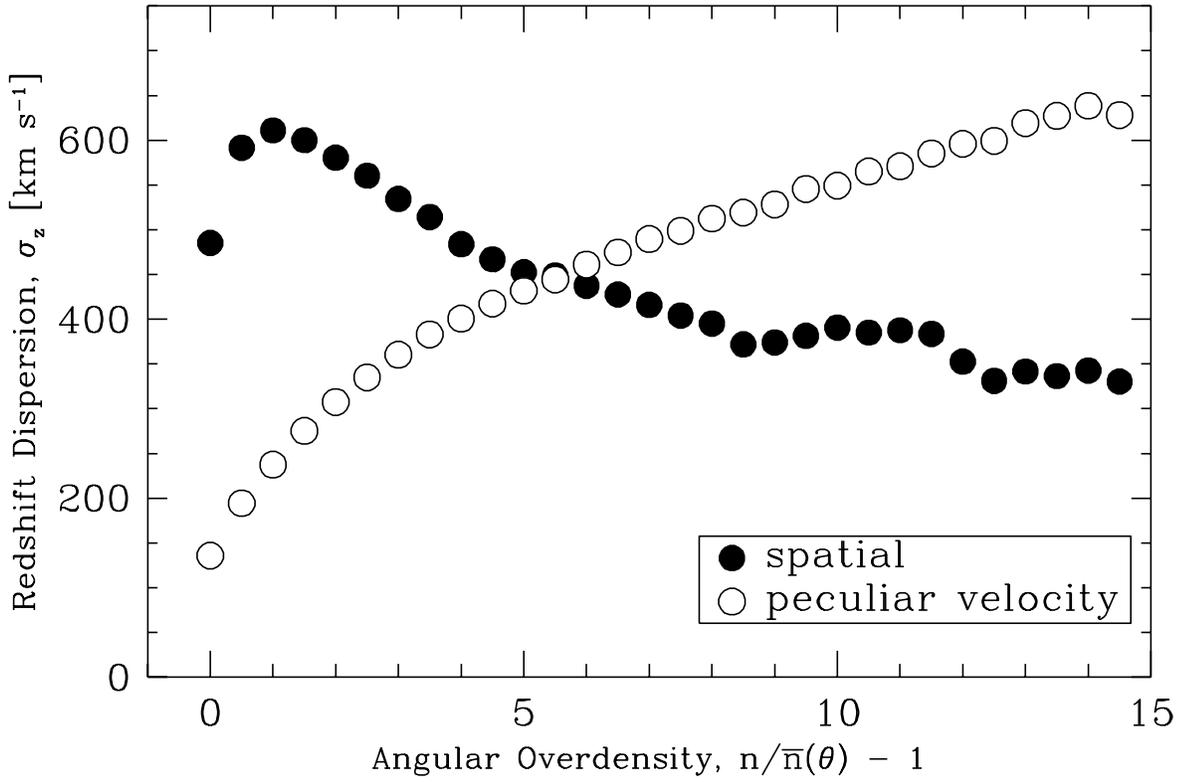}{10cm}{0}{90}{90}{-300}{-350}
\caption{
  Components of the redshift dispersion, $\sz$, as a function of angular 
overdensity for $\Omega=1$ CDM. The spatial component represents the
dispersion due to the positions of the particles, while the peculiar
velocity component represents the dispersion from the motions of the
particles themselves. The measured redshift dispersion is a combination
of both components (see Figure~5).
}\end{figure}

\begin{figure}
\plotfiddle{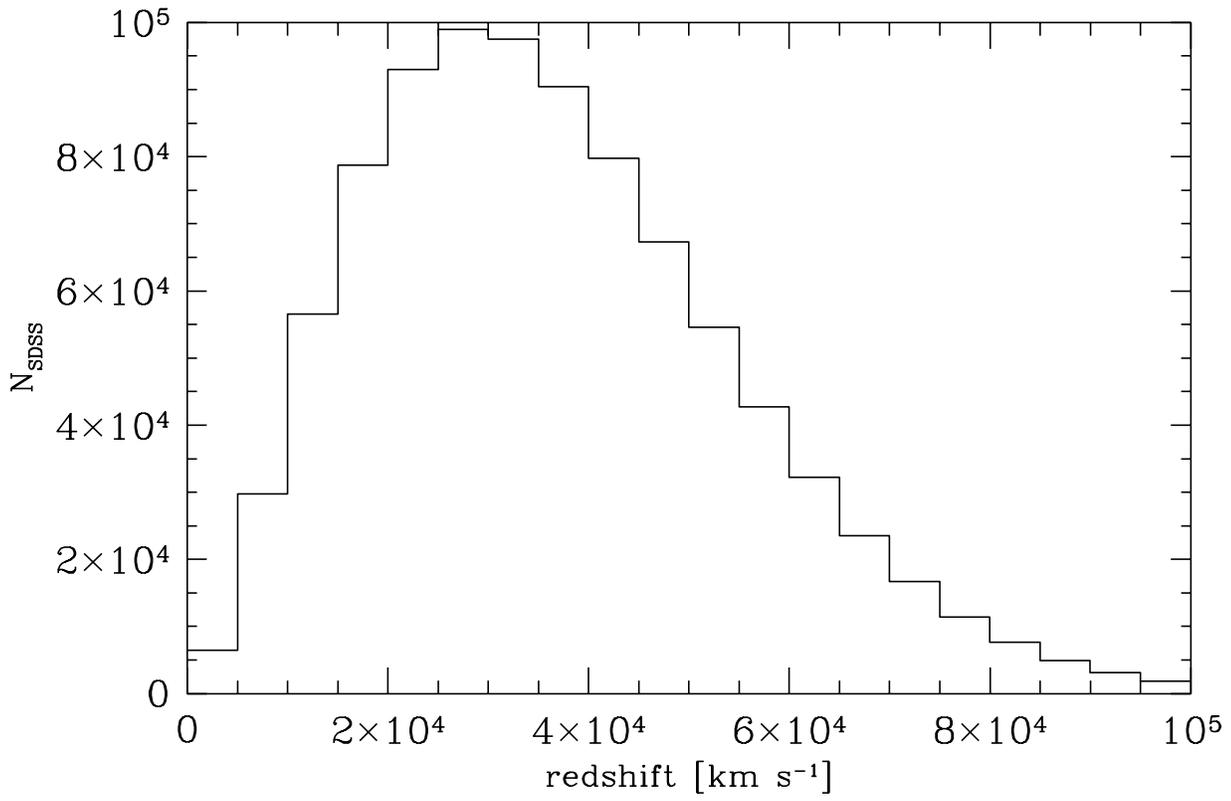}{10cm}{0}{90}{90}{-300}{-350}
\caption{
  Estimated number of galaxies in each volume limited shell of the
SDSS.
}
\end{figure}

\begin{figure}
\plotfiddle{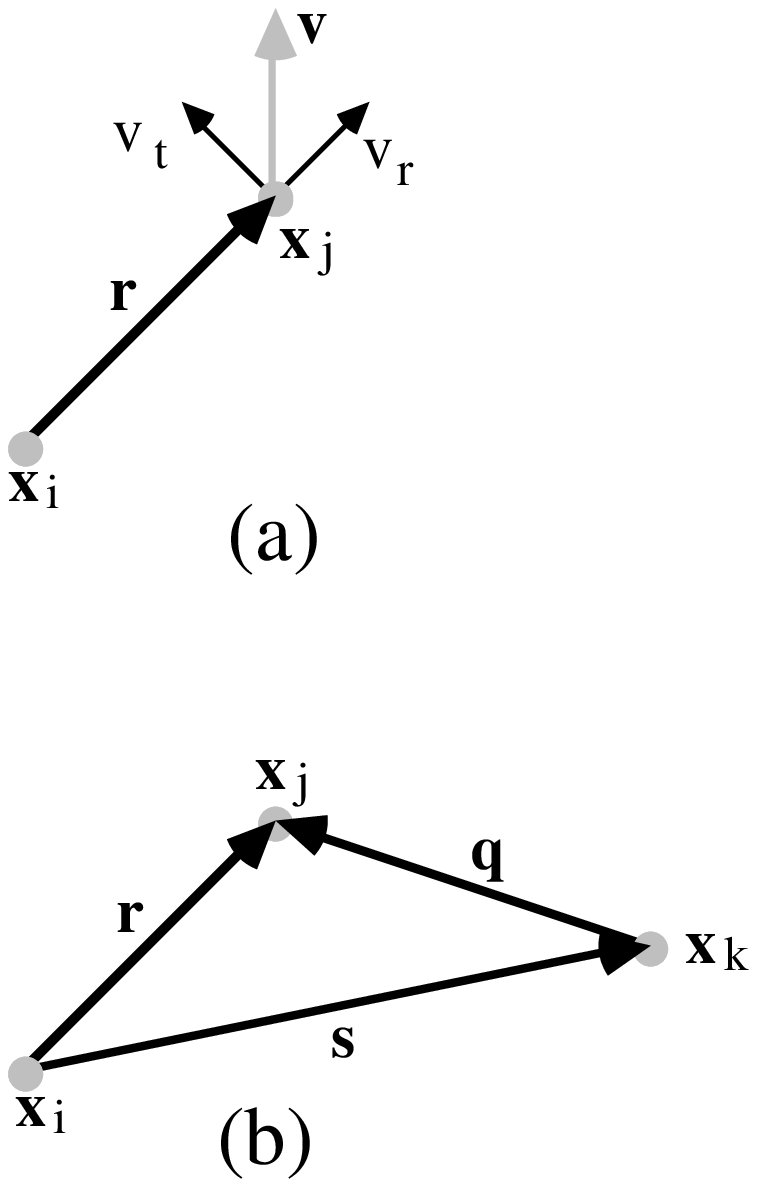}{16cm}{0}{150}{150}{-265}{-550}
\caption{
  Schematic drawings depicting positions and velocities  of {\bf (a)}
two particle and {\bf (b)} three particle systems  used to derive the
Cosmic Virial Theorem.
}
\end{figure}

%%%%%%%%%%%%%%%%%%%%%%%%%%%%%%%%%%%%%%%%%%%%%%%%%%%%%%%%%%%%

\end{document}